\documentclass[aps,prl,reprint,superscriptaddress,longbibliography,nobibnotes]{revtex4-2}
\usepackage{amsmath,amssymb}
\usepackage{graphicx}
\usepackage{bm}
\usepackage{siunitx}

\usepackage[normalem]{ulem}
\def\url#1{}

\begin{document}

\newcommand{\Sqw}{${\cal S}(\tilde{q},\omega)$}
\newcommand{\Sqwcal}{${\cal S_{\rm cal}}(\tilde{q},\omega)$}

\title{Quasiparticle Interactions in the $S = 1/2$ Heisenberg Antiferromagnetic Chain CuPzN}

\author{Leshan Zhao}
    \email{lzhao53@jhu.edu}
    \affiliation{William H. Miller III Department of Physics \& Astronomy, Johns Hopkins University, Baltimore, Maryland 21218, USA}
    \author{Timothy Reeder}
    \affiliation{William H. Miller III Department of Physics \& Astronomy, Johns Hopkins University, Baltimore, Maryland 21218, USA}
    \affiliation{Condensed Matter Physics and Materials Science Department, Brookhaven National Laboratory, Upton, NY 11973, USA}
    \author{Mathew Stone}
    \affiliation{Neutron Scattering Division, Oak Ridge National Laboratory, Oak Ridge, Tennessee 37831, USA}
    \author{Daniel G. Mazzone}
    \author{Jakob Lass}
    \author{Christof Niedermayer}
    \affiliation{PSI Center for Neutron and Muon Sciences, 5232 Villigen PSI, Switzerland}
    \author{Daniel H. Reich}
    \affiliation{William H. Miller III Department of Physics \& Astronomy, Johns Hopkins University, Baltimore, Maryland 21218, USA}
    \author{Anna Keselman}
    \affiliation{Department of Physics, Technion, Haifa, 3200003, Israel}
    \affiliation{The Helen Diller Quantum Center, Technion, Haifa, 3200003, Israel}
    \author{Oleg A. Starykh}
    \affiliation{Department of Physics and Astronomy, University of Utah, Salt Lake City, Utah 84112, USA}
    \author{Collin Broholm}
    \email{broholm@jhu.edu}
    \affiliation{William H. Miller III Department of Physics \& Astronomy, Johns Hopkins University, Baltimore, Maryland 21218, USA}

\date{\today}

\begin{abstract}
We investigate spinon--spinon interactions in the quasi-one-dimensional
$S=1/2$ Heisenberg antiferromagnet Cu(C$_4$H$_4$N$_2$)(NO$_3$)$_2$ (CuPzN) using inelastic neutron scattering. At zero field, the spectrum exhibits the canonical multi-spinon continuum of the Heisenberg chain. With increasing field we demonstrate the appearance of an energy gap $2\Delta$ towards momentum $q\rightarrow 0$ between the uniform magnetization precession mode and the shifted spinon continuum. $2\Delta$ is a direct measure of spinon backscattering interactions. Comparison with numerical matrix-product-state calculations and a recent analytical theory shows excellent agreement. By examining the integrable spin-1/2 chain, we provide an example of how quasiparticle interactions can be quantified experimentally in isotropic quantum spin systems.
\end{abstract}

\maketitle

\textit{Introduction--} Quantum spin liquids (QSL) are a distinct state of matter defined by quantum entanglement that can support exotic fractionalized quasi-particles~\cite{Savary_2017,doi:10.1126/science.aay0668}. Theoretical work has established QSL phases in idealized models, but identifying their signatures unambiguously in real materials remains a major challenge. 

A hallmark of QSLs is the fractionalization of spin-1 magnons into spin-1/2 spinons~\cite{Savary_2017,knolle_moessner_2019,doi:10.1126/science.aay0668}. While neutron scattering provides access to dynamic correlations between spins~\cite{boothroyd2020principles}, fractionalization results in continuum scattering, the analysis of which requires detailed knowledge of the spin Hamiltonian. Lacking this, there is a need for model-independent analysis of the scattering data. Quantum Fisher information extracted from neutron scattering, for example, has been shown to provide a lower bound on entanglement~\cite{https://doi.org/10.1002/qute.202400196,hauke_heyl_tagliacozzo_zoller_2016}. Separately there are proposals to use path-spin entangled neutron beams to directly probe entanglement in QSLs~\cite{Irfan_2021}. 

Here we demonstrate a symmetry-based experimental method to determine residual interactions between spinons in a spin-conserving QSL. An external magnetic field tunes the spinon chemical potential and modifies the kinematics of the spinon continuum, imprinting signatures of spinon--spinon interactions in the dynamical structure factor ${\cal S}(\mathbf{q},\omega)$~\cite{cauxmaillet2005,pereira2006,stone2003,kuhne2011}. Specifically, the magnetic field redistributes the spin spectral weight by shifting the spinon continuum to higher energies near $\mathbf{q} \rightarrow 0$ while simultaneously inducing a spinon collective mode that emerges exactly at the Zeeman energy~\cite{keselman_balents_starykh_2020}. The resulting energy gap between the continuum and the collective mode is directly accessible through neutron scattering and provides a quantitative measure of spinon interactions. Experimentally establishing these signatures provides a test of the quasiparticle description of correlated quantum matter and sensitivity to sub-leading interactions that can otherwise be difficult to resolve.

To test this method, we focus on the nearest-neighbor spin-$1/2$ Heisenberg antiferromagnetic chain where quantum fluctuations suppress conventional long-range N\'eel order~\cite{mermin1966}. The experimental realization of this model provides a unique opportunity for quantitative comparisons between experiment and theoretical predictions including non-perturbative theory enabled by the exact solubility of this model~\cite{bethe1931,cauxhagemans2006,kohno2009}, essentially exact numerical treatments by matrix-product-state (MPS) calculations, and field theory methods~\cite{giamarchi2003quantum}.

The spin-$1/2$ Heisenberg chain provides a paradigmatic example of quasiparticle fractionalization of spin-$1$ excitations into pairs of spin-$1/2$ spinons~\cite{faddeev1981,cauxmaillet2005,pereira2006}. The resulting multi-spinon continuum is an experimental signature of quasiparticle fractionalization that is well established in inelastic neutron scattering studies of spin-chain materials~\cite{tennant1995,stone2003,dallapiazza2015,Mourigalchain}. 

Copper pyrazine dinitrate, Cu(C$_4$H$_4$N$_2$)(NO$_3$)$_2$ (CuPzN) is an excellent material platform for our purposes~\cite{losee1973,hammar1999,stone2003,kuhne2011}. The dominant magnetic super-exchange runs along chains of Cu$^{2+}$ ions linked by pyrazine molecules with a nearest-neighbor exchange of $J=\SI{0.90(1)}{meV}$. 3D magnetic ordering is absent down to $T_N=\SI{107}{mK}$, indicating very weak inter-chain interactions ($J'/J<10^{-4}$)~\cite{hammar1999,stone2003}. 

In this Letter, we use neutron scattering from magnetized CuPzN to examine the effects of spinon interactions in the dynamical structure factor ${\cal S}(\mathbf{q},\omega)$. In zero field, the scattering is consistent with the exact multi-spinon contribution to ${\cal S}(\mathbf{q},\omega)$. In the magnetized state, an energy gap $2\Delta$ develops between the collective spinon branches for $\mathbf{q} \rightarrow 0$, providing an experimental measure of spinon interactions. The extracted spinon energy gap and the dispersion and spectral-weight are in quantitative agreement with matrix-product state (MPS) calculations and analytical theory~\cite{keselman_balents_starykh_2020}. 

To connect the experimentally measured energy gap $2\Delta$ to spinon interactions, we summarize an analytical description of the interacting spinon modes and the physical origin of their field-induced splitting~\cite{keselman_balents_starykh_2020}. We then compare our experimental results with the analytical description and numerical calculations, providing a stringent test of the theory and its experimental relevance. 

\textit{Theoretical Results--} The finite-magnetization response function of the Heisenberg chain can be described analytically in terms of two collective modes whose energies and spectral weights are modified by spinon backscattering~\cite{keselman_balents_starykh_2020}. The corresponding dynamical structure factor is given by ${\cal S}(\tilde{q},\omega) = \pi^{-1} (1 - e^{-\hbar \omega/k_B T})^{-1} {\rm Im}\,\chi(\tilde{q},\omega)$, where
\begin{eqnarray}
\label{eq:susceptibility}
\chi^{\pm}(\tilde{q},\omega) &=& \frac{M}{\hbar}\left(\frac{A_{+}(\tilde{q})}{\omega-\omega_{+}(\tilde{q})} + \frac{A_{-}(\tilde{q})}{\omega-\omega_{-}(\tilde{q})}\right), \\
\label{eq:spectral_weight}
A_{\pm}(\tilde{q}) &=& 1 \pm \frac{\hbar^2 v^{2}\tilde{q}^{2}-B\Delta}{B\sqrt{\Delta^{2}+\hbar^2v^2 \tilde{q}^{2}}}, \\
\hbar \omega_{\pm}(\tilde{q}) &=& B + \Delta \pm \sqrt{\Delta^{2}+\hbar^2 v^2\tilde{q}^{2}}.
\label{eq:dispersion}
\end{eqnarray}
Here $M=\langle S^z\rangle$ is the dimensionless equilibrium magnetization per spin, $B=g_e\mu_B\mu_0H$ denotes the Zeeman energy with $\mu_0H$ being the applied magnetic flux density, $\tilde{q} = q a$ is the dimensionless transferred momentum along the chain direction, $v=v_0\sqrt{1-\delta^2}$ is the renormalized spinon velocity in which $v_0=\pi J/2\hbar$ is the spinon Fermi velocity, and $g_e$ is the electronic Land\'e factor. The dimensionless parameter $\delta= g J/(4\pi \hbar v_0)$ describes the strength of the back-scattering interaction $g$ between spinons. It determines the $\tilde{q}=0$ splitting between $\omega_\pm$ branches via $2\Delta = g J M$~\cite{keselman_balents_starykh_2020}. This definition makes $g$ dimensionless. Alternatively, one can describe the physics using the spinon back-scattering velocity $u = g J/\hbar$, so that $\delta = u/(4 \pi v_0)$ and $2\Delta = \hbar u M$. 

The physical meaning of this description can be understood considering the equations of motion for the operators of magnetization $\hat{\bf m}(x,t)$ and magnetization current $\hat{\bf j}(x,t)$, derived in Ref.~\cite{rbwang2022}. The zero-momentum $\tilde{q}=0$ components of these operators satisfy 
\begin{equation}
        \partial_t \hat{\bf m}_0  = - \frac{1}{\hbar} {\bf B} \times \hat{\bf m}_0, 
        \, \partial_t \hat{\bf j}_0  = -  \frac{1}{\hbar}({\bf B} + Jg {\bf M}) \times \hat{\bf j}_0, 
        \label{eq:larmor}
\end{equation}
where, as explained above, ${\bf M}$ is the equilibrium magnetization along the external magnetic field ${\bf H}$, and $g M = 2\Delta / J$. These expressions show that the $\tilde{q}=0$ component of $\hat{\bf m}(x,t)$ precesses at exactly the Zeeman frequency $\hbar\omega_{-}(\tilde{q}=0)=B$, as required by the Larmor theorem for spin-rotationally invariant systems. The magnetization current $\hat{\bf j}_0$ is, however, not protected by this symmetry and experiences the total internal field of magnitude $\hbar\omega_{+}(\tilde{q}=0) = B + Jg M=B+2\Delta$, the second term of which represents the ``molecular'' field of spinons. At finite momentum, the two excitation branches hybridize and repel, resulting in opposite $\pm$ signs in Eq.~\eqref{eq:dispersion}.

\begin{figure}[t]
    \centering
    \includegraphics[width=\linewidth]{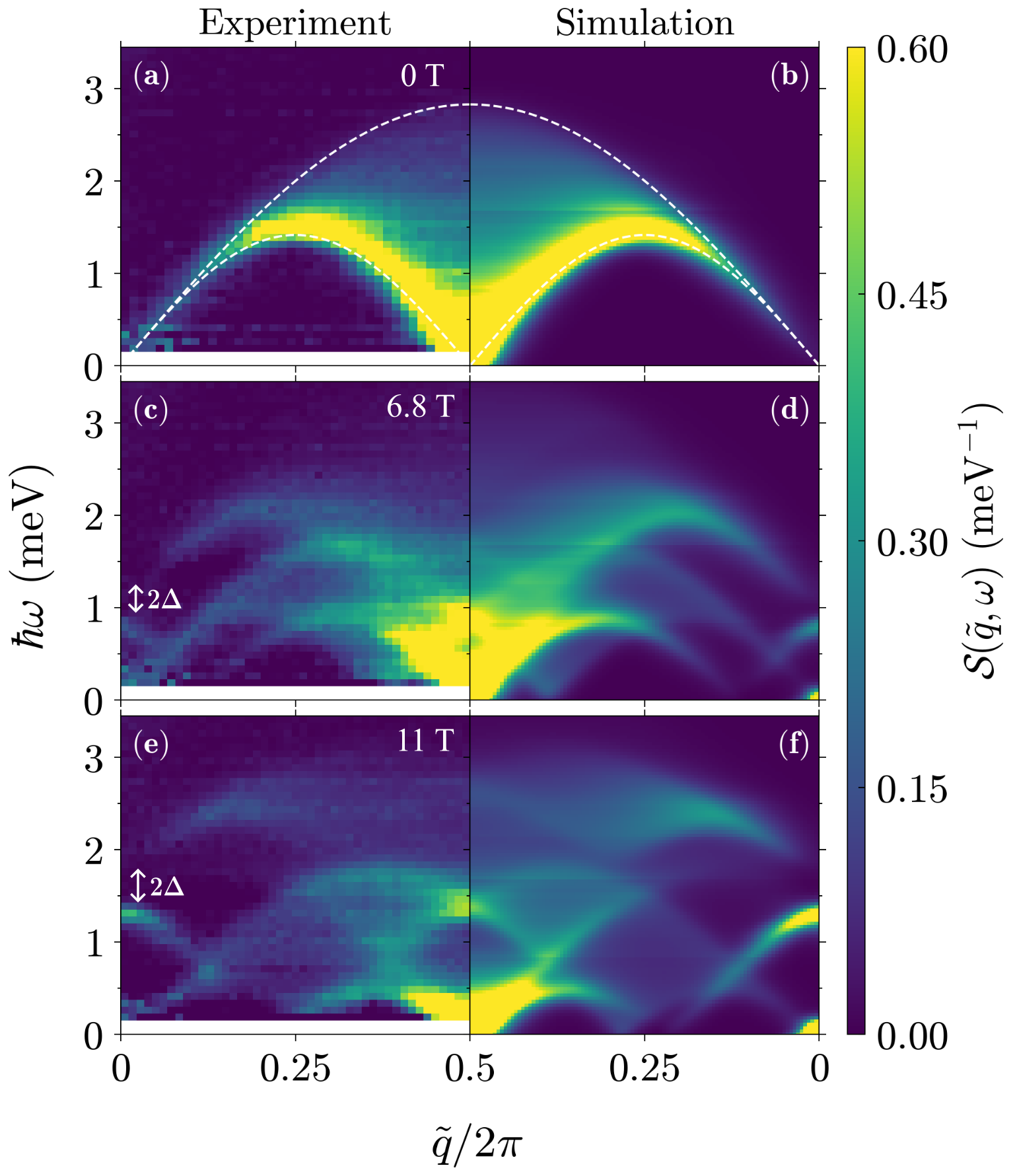}
    \caption{
    Dynamical structure factor \Sqw\ of CuPzN measured by neutron scattering at $T=\SI{50}{mK}$ in  magnetic fields (a) $\mu_0 H=\num{0}$, (c) $\mu_0 H=\SI{6.8}{T}$, and (e) $\mu_0 H=\SI{11}{T}$ applied along the {\bf c}-axis, which is perpendicular to the {\bf a}-oriented spin chain. The $2\Delta$ energy gap between the first and second excited states near $\tilde{q} \rightarrow0$ is indicated in (c) and (e). White dashed lines in (a) indicate the two-spinon continuum boundaries calculated for a nearest-neighbor exchange interaction $J=\SI{0.9}{meV}$. (b, d, f) Corresponding dynamical structure factor \Sqwcal, calculated using time-evolution block decimation (TEBD) from a matrix-product state (MPS) ground state obtained by the density-matrix renormalization group (DMRG), corrected for instrumental resolution.}
    \label{fig:CuPzN_QE}  
\end{figure}

\textit{Experimental Results--} To test the experimental relevance of these results, inelastic neutron scattering measurements were performed on a co-aligned mosaic of CuPzN single crystals in the $(hk0)$ scattering plane with a magnetic field ${\bf H} \parallel {\bf c}$ at the multiplexing spectrometer CAMEA at the Swiss Spallation Neutron Source SINQ, Paul Scherrer Institute~\cite{10.1063/5.0128226}. The sample contained 22 single crystals of deuterated CuPzN with a total mass of 3.1 g, and a full width at half maximum mosaic of 2$\rm ^o$ for rotation about $\bf c$. The data were processed with the MJOLNIR software package~\cite{ lass_jacobsen_mazzone_lefmann_2020}. We combine data collected with incident neutron energies $E_i=\SI{3.8}{meV}$, \SI{5.2}{meV}, and \SI{6.9}{meV}. For energy transfers $\hbar\omega<\SI{0.7}{meV}$, only $E_i=\SI{3.8}{meV}$ data were used, bringing the energy resolution to $\delta E=\SI{0.1}{meV}$ at the elastic line. We obtained the isotropic part of the non-magnetic background by taking a powder average of the zero-field detector count rate at each momentum transfer $|\mathbf{q}|$ that lies at least \SI{0.2}{meV} outside the known bounds of the zero-field two-spinon continuum for CuPzN. The resulting background profile $I_{\mathrm{BG}}(|\mathbf{q}|,\hbar\omega)$ was subtracted from the raw data at all magnetic fields. The form-factor-corrected magnetic scattering $I_m({\bf q},\hbar\omega)/|F(|{\bf q}|)|^2$ was then projected onto the first Brillouin zone of the chain direction using $\tilde{q}={\bf q}\cdot {\bf a}$ and normalized using the total moment sum-rule. In the vertical field configuration used here, we measure the average of the longitudinal and the transverse correlations: ${\cal S}(\tilde{q},\omega)=\frac{1}{2}({\cal S}^{\perp}+{\cal S}^{zz})$ where ${\cal S}^{\perp} \equiv (S^{xx}+S^{yy})/2$ and we neglect magnetic anisotropy in the plane perpendicular to $\bf H$.

The results shown in Fig.~\ref{fig:CuPzN_QE} provide a complete picture of the field dependent multi-spinon continuum in CuPzN. Fig.~\ref{fig:CuPzN_QE}(a) shows the experimental zero-field continuum scattering. The dashed lines show the exact lower bound $\hbar\omega_L=(\pi J/2)|\sin\tilde{q}|$ and the approximate upper bound $\hbar\omega_U=\pi J|\sin(\tilde{q}/2)|$ of the two-spinon continuum for $J=\SI{0.9}{meV}$~\cite{faddeev1981}. These lines enclose most of the spectral weight from 2- and 4-spinon excitations~\cite{cauxhagemans2006}. Fig.~\ref{fig:CuPzN_QE}(b) shows the dynamical structure factor ${\cal S}_{\rm cal}(\tilde{q},\omega)$ calculated with time-evolution block decimation (TEBD) from a matrix-product-state (MPS) ground state obtained by density-matrix renormalization group (DMRG)~\cite{keselman_balents_starykh_2020,white1992,PhysRevLett.93.040502}. The numerical calculations were performed using the ITensor library~\cite{itensor} on systems of $N=200$ sites, with the time evolution carried out up to $t_{\rm max}=30\hbar J^{-1}$. The calculated spectra are shown after convolution with the resolution function of the experimental instrument. A pixel-by-pixel fit of the zero-field ${\cal S}_{\rm cal}(\tilde{q},\omega)$ to ${\cal S}(\tilde{q},\omega)$ yields $J=\SI{0.90(1)}{meV}$, in agreement with previous reports~\cite{hammar1999,stone2003}.

\begin{figure}[t]
    \centering
    \includegraphics[width=\linewidth]{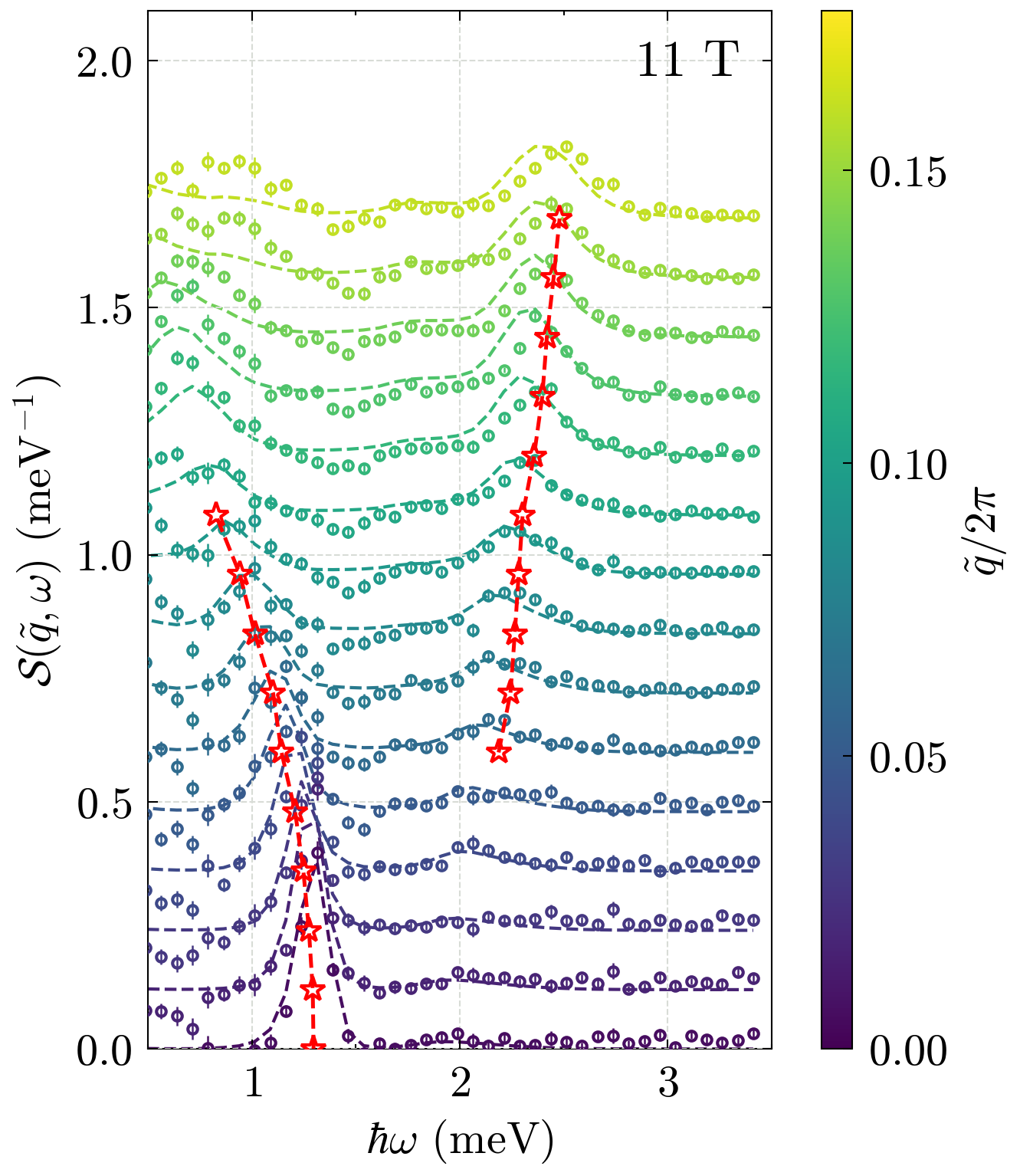}
    \caption{Dynamical structure factor \Sqw\ as a function of energy transfer $\hbar\omega$ at fixed momenta $\tilde{q}$ along the spin chain for CuPzN, measured at $\mu_0H=\SI{11}{T}$. Curves at different $\tilde{q}$ are vertically offset by \SI{0.015}{meV^{-1}} for clarity. Dashed lines show \Sqwcal\ calculated numerically, as described in text. Red stars indicate the centers of spinon dispersion peaks, extracted from Gaussian fits to the measurements. Red lines show dispersion relations determined from the peak positions.}
\label{fig:CuPzN11T_QE_cuts}
\end{figure}

\begin{figure}[t]
    \centering
    \includegraphics[width=\linewidth]{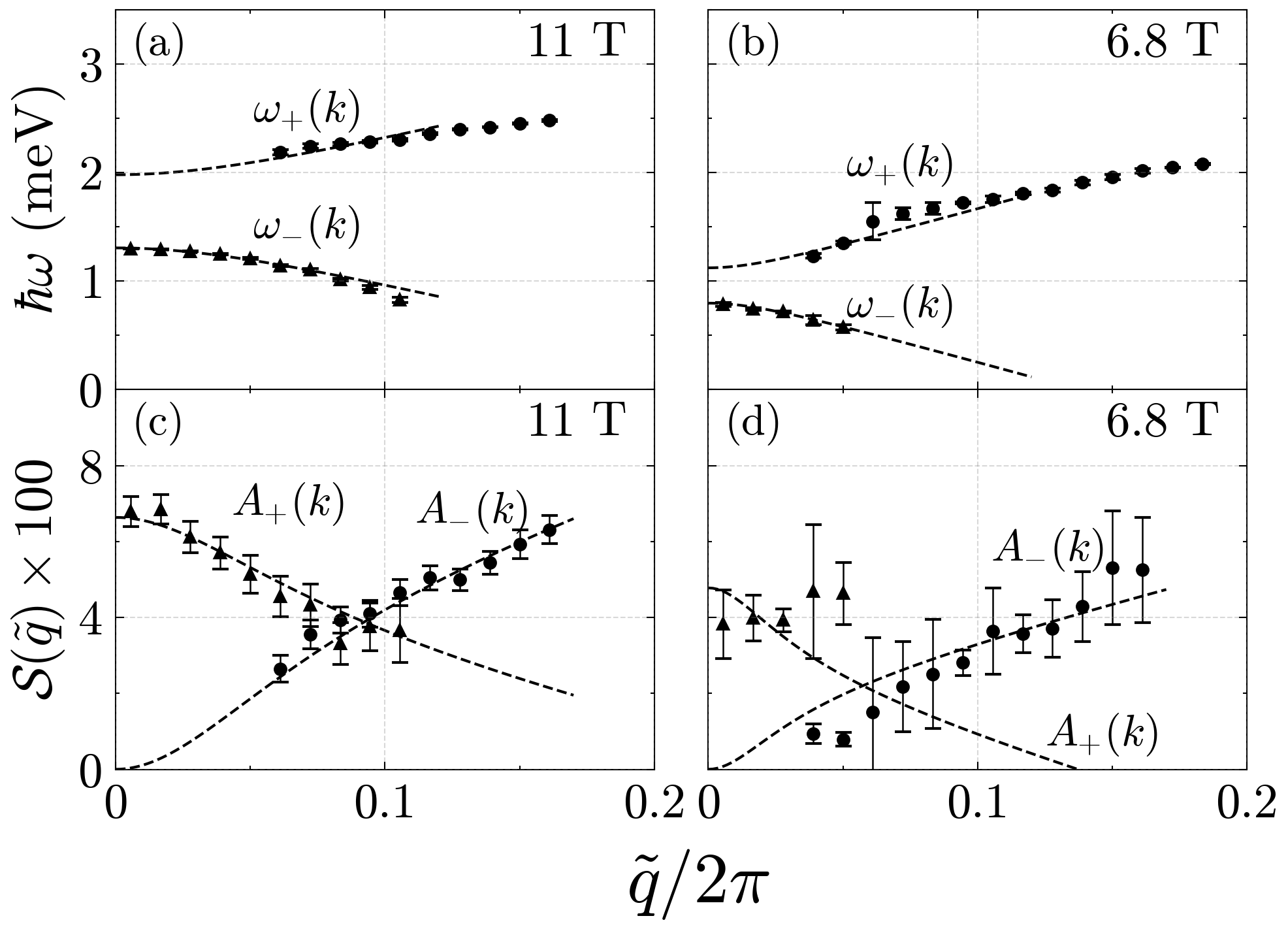}
    \caption{Dispersion relations $\hbar\omega_\pm(q)$ for collective spinon modes in CuPzN measured at (a) $\mu_0H=\SI{11}{T}$ and (b) $\mu_0H=\SI{6.8}{T}$, extracted from Gaussian fits to constant-$\tilde{q}$ cuts through \Sqw. Dashed lines show fits to Eq.~\eqref{eq:dispersion}, from which the energy gap $2\Delta$ at $\tilde{q}=0$ was extracted. (c, d) Spectral weights ${\cal S}_\pm(\tilde{q})$ for the collective spinon modes measured at (c) $\mu_0H=\SI{11}{T}$ and (d) $\mu_0H=\SI{6.8}{T}$. Dashed lines represent the spectral weights calculated using Eq.~\eqref{eq:spectral_weight} with parameters listed in Table \ref{tab:fit-results}.}
    \label{fig:11T_Peaks}
\end{figure}

We next study the spectrum of CuPzN in applied magnetic fields up to approximately $1/2$ of the saturation magnetization. At $\mu_0H=\SI{6.8}{T}$ where $M=\langle S^z \rangle =\num{0.12(1)}$~\cite{hammar1999}, a finite energy gap $2\Delta$ emerges between the first and second excited states near $\tilde{q}\rightarrow 0$, as shown in Fig.~\ref{fig:CuPzN_QE}(c). Upon increasing the field to $\mu_0H=\SI{11}{T}$ where $M=\num{0.24(1)}$, $2\Delta$ increases further (Fig.~\ref{fig:CuPzN_QE}(e)). Fig.~\ref{fig:CuPzN11T_QE_cuts} shows constant $\tilde{q}$ cuts through ${\cal S}(\tilde{q} ,\hbar\omega)$ at $\mu_0H=\SI{11}{T}$, with dashed lines representing ${\cal S_{\rm cal}}(\tilde{q} ,\hbar\omega)$ at each $\tilde{q}$. Stars in red are centers of resonant magnetic excitations $\hbar\omega_{\pm}(\tilde{q})$ at each $\tilde{q}$, determined by Gaussian fits to the data. Since the spectral weight of the second excited state vanishes as $\tilde{q}\rightarrow 0$, as expected from Eq.~\eqref{eq:spectral_weight}, the energy gap $2\Delta$ at $\tilde{q}=0$ is not directly measurable in the data. Instead, we determine $2\Delta$ by fitting the measured $\hbar\omega_{\pm}(\tilde{q})$ to the dispersion relation in Eq.~\eqref{eq:dispersion}. Results of the dispersion fits for these two applied fields are in Fig.~\ref{fig:11T_Peaks}(a) and (b), with parameters summarized in Table~\ref{tab:fit-results}. This analysis yields $g=2\Delta/MJ=3.0(3)$ and $\num{3.1(5)}$ for $\mu_0H=\SI{6.8}{T}$ and $\mu_0H=\SI{11}{T}$, respectively.

Figure~\ref{fig:11T_Peaks}(c) and (d) show the spectral weights ${\cal S}(\tilde{q}) = \int {\cal S}(\tilde{q}, \omega)d\hbar\omega$ of the two spinon dispersion branches near $\tilde{q} \rightarrow 0$, with dashed lines representing Eq.~\eqref{eq:spectral_weight} with the parameters from Table~\ref{tab:fit-results}. Excellent agreement was found between the measured ${\cal S}(\tilde{q})$ and the calculated $A_{\pm}(\tilde{q})$ with parameters obtained from the dispersion fits described above.

\begin{figure}[t]
    \centering
    \includegraphics[width=\linewidth]{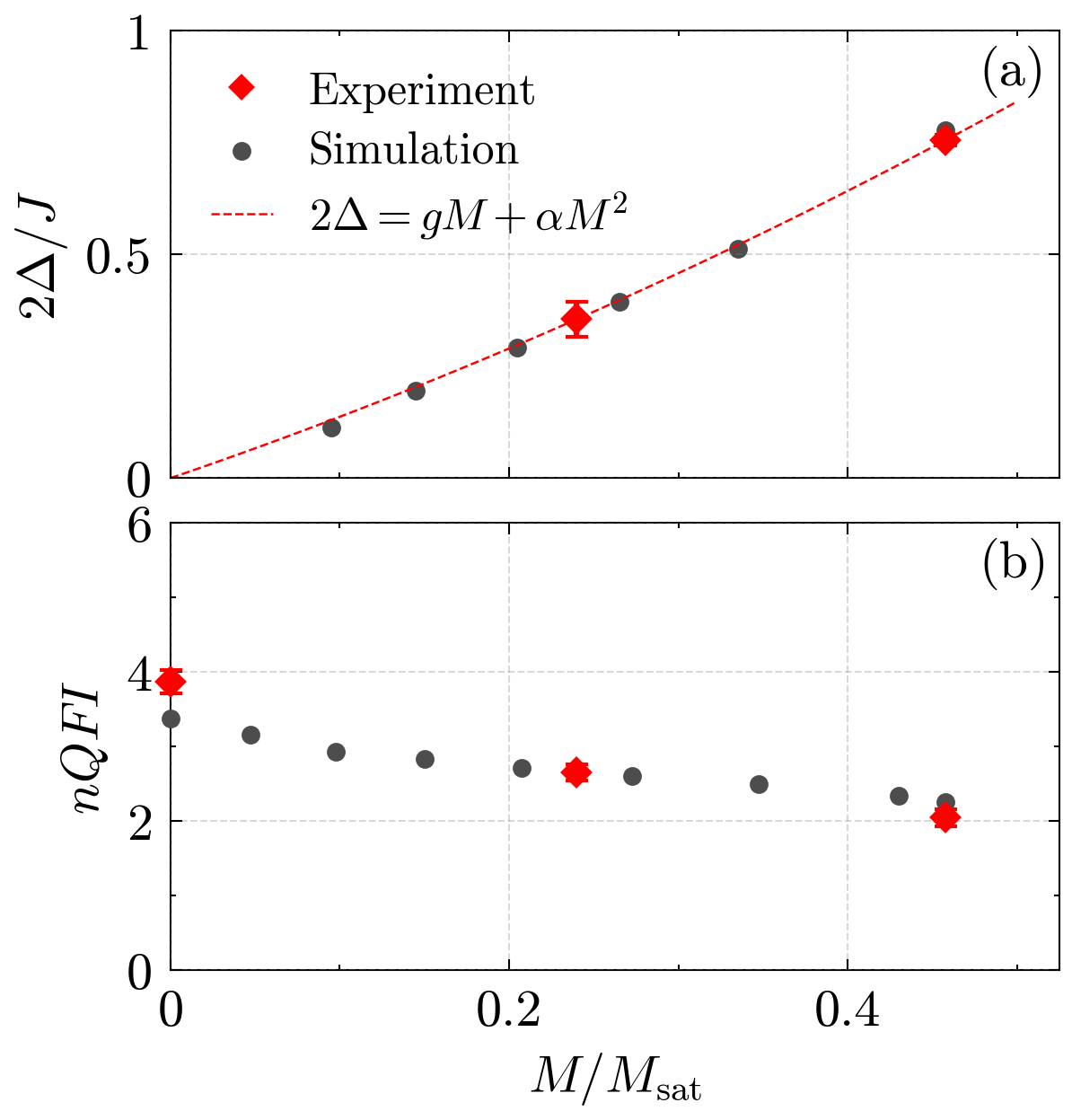}
    \caption{(a) Experimental (red) and calculated (black) energy gap $2\Delta$ between the uniform magnetization mode and the shifted spinon continuum at $\tilde{q}=0$ versus the relative magnetization $M/M_{\rm sat}$, where $M_{\rm sat}=1/2$ is the dimensionless saturation magnetization. The dashed line shows a quadratic fit to the measured data. The linear coefficient $g$ represents contribution from two-spinon back-scattering interaction, and $\alpha$ accounts for higher order effects not captured in the analytical theory~\cite{keselman_balents_starykh_2020}. (b) Normalized quantum Fisher information (nQFI) for CuPzN versus $M/M_{\rm sat}$, with red dots denoting experimental measurements and black dots numerical calculations convolved with instrumental resolution.
    }
    \label{fig:DMRG_comparison}
    
\end{figure}

Figure~\ref{fig:DMRG_comparison}(a) shows the $\tilde{q}=0$ gap $2\Delta$ between the two lowest energy modes as a function of the relative magnetization $M/M_{\rm sat}=2M$ and compares it with numerical results for the nearest-neighbor Heisenberg antiferromagnetic chain. The numerical $2\Delta$ was obtained following a similar fitting procedure from constant-$\tilde{q}$ cuts of \Sqwcal\ obtained from MPS calculations. To obtain the best estimate of $g$ from both experimental datasets, we fit $2\Delta(M)$ to the quadratic form $2\Delta/J=gM+\alpha M^2$. The fit yields a spinon backscattering strength of $g=\num{2.3(4)}$ and a quadratic coefficient $\alpha=\num{3(2)}$, indicating the presence of higher-order corrections beyond those retained in the low-energy approximation used to derive Eqs.~(1--3)~\cite{keselman_balents_starykh_2020}. The agreement with the numerical simulation is excellent. This value of $g=2.3(4)$ as the quantitative experimental measure of spinon-spinon interactions in CuPzN corresponds to a back-scattering energy scale of $gJ =\SI{2.1(4)}{meV}$. The backscattering interaction is strong compared to the magnetic exchange interaction. Yet the dimensionless parameter controlling the theoretical calculation is quite small, $\delta \approx 0.12$, and the theory~\cite{keselman_balents_starykh_2020,rbwang2022} is internally consistent. In terms of the spinon backscattering velocity, $u/v_0 \approx 1.46$. 

\begin{table}
\caption{Parameters in $\hbar\omega_\pm(\tilde{q})$ of Eq.~\eqref{eq:dispersion} obtained by fitting peak positions in Fig.~\ref{fig:CuPzN_QE}(c, e) for $\tilde{q}\rightarrow 0$, as illustrated in Fig.~\ref{fig:CuPzN11T_QE_cuts}.}
\label{tab:fit-results}
\begin{ruledtabular}
\begin{tabular}{ccc}
 & $\mu_0H=\SI{6.8}{T}$ & $\mu_0H=\SI{11}{T}$ \\
\hline
$B$ (meV) & 0.80(1) & 1.305(2) \\
$2\Delta$ (meV) & 0.32(4) & 0.68(2) \\
$\hbar v$ (meV) & 1.09(5) & 0.94(1) \\
$g$ & 3.0(3) & 3.1(5) \\
\end{tabular}
\end{ruledtabular}
\end{table}

\begin{figure}[t]
    \centering
    \includegraphics[width=\linewidth]{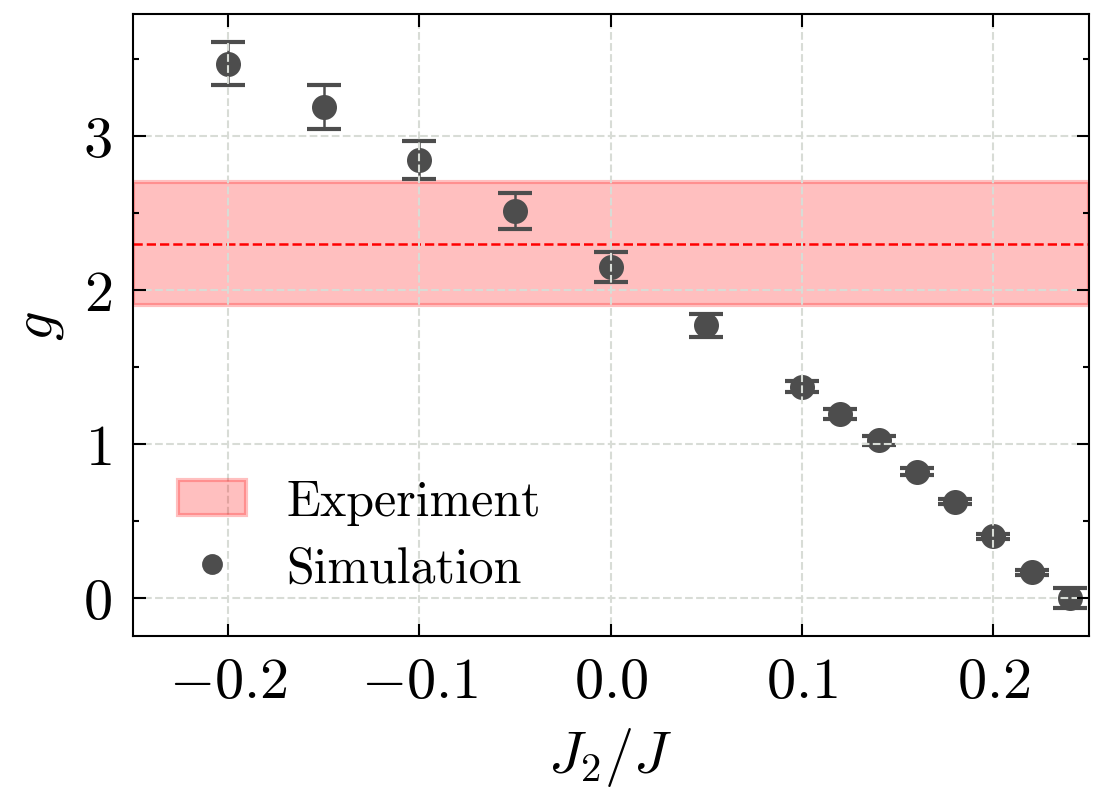}
    \caption{Two-spinon back-scattering interaction strength $g$ as a function of the second-nearest-neighbor exchange interaction $J_2/J$. Black points show numerically calculated values, and the red shaded region indicates the range within one standard deviation of the experimental measurement. A constraint on the second-nearest-neighbor exchange in CuPzN of $-0.04 <J_2/J< 0.02$ is found.}
    \label{fig:g_vs_J2}
\end{figure}

The spinon back-scattering interaction $g$ is sensitive to perturbations beyond the nearest-neighbor interaction \cite{keselman_balents_starykh_2020}, and our measurement of $g$ for CuPzN can be used to constrain sub-leading interactions. Figure~\ref{fig:g_vs_J2} compares the experimentally determined $g$ with numerical calculations for varying second-nearest-neighbor interaction $J_2$. This comparison puts a constraint on the second-nearest-neighbor exchange in CuPzN of $-0.04 <J_2/J< 0.02$.

From the dynamical structure factor \Sqw\ at $\tilde{q}=\pi$, the normalized quantum Fisher information (nQFI),
\begin{equation*}
\begin{aligned}
\mathrm{nQFI} ={}& 4\int_0^{\infty} d(\hbar\omega)\,
\tanh\!\left(\frac{\hbar\omega}{2k_BT}\right)
\left(1-e^{-\hbar\omega/k_BT}\right) \\
&\times {\cal S}(\tilde{q}=\pi,\hbar\omega),
\end{aligned}
\end{equation*}
provides a lower bound on multipartite entanglement~\cite{hauke_heyl_tagliacozzo_zoller_2016, scheie_laurell_simeth_dagotto_tennant_2025}. Fig.~\ref{fig:DMRG_comparison}(b) shows nQFI as a function of the relative magnetization $M/M_{\rm sat}$. The experimental measurements and the numerical calculations show a consistent decrease in nQFI with increasing magnetization. This reflects the redistribution of spectral weight away from the antiferromagnetic wavevector $\tilde{q}=\pi$ and a reduction in multipartite entanglement as the spin chain is driven towards the unentangled ferromagnetic state. 

\textit{Conclusion--} We have used inelastic neutron scattering to resolve signatures of spinon--spinon interactions in the dynamical structure factor of CuPzN, a near-ideal realization of the $S=1/2$ Heisenberg antiferromagnetic chain. At zero field, the measured spectrum is consistent with the canonical multi-spinon continuum. At finite field, the $2\Delta$ energy splitting between low-$q$ collective spinon excitations provides a direct experimental measure of spinon backscattering. The experimentally determined energy gap, dispersion, and spectral-weight evolution agree quantitatively with MPS calculations and the analytical theory accounts for the $M\rightarrow 0$ limit~\cite{keselman_balents_starykh_2020,rbwang2022}.

Our findings of the interaction-induced spectral gap agrees with an earlier electron spin resonance-based confirmation of the interacting spinon picture in the quasi-one-dimensional spin-chain material K$_2$CuSO$_4$Br$_2$~\cite{povarov2022}. Our results for the small-$q$ spinon dispersion $\omega_\pm(\tilde{q})$ and the corresponding spectral weights $A_\pm$ represent the first experimental determination of these key parameters of the interacting spinon liquid. 

Importantly, these phenomena are not restricted to one-dimensional quantum magnets. The magnetic-field-induced shift of the spinon continuum above
the Zeeman frequency and the emergence of the interaction-induced collective spinon mode 
are generic features of spin-rotation-invariant quantum spin liquids hosting fractionalized spinon excitations and emergent gauge fields~\cite{balents2020}. Our work thus demonstrates a new tool to characterize candidate quantum spin liquid materials~\cite{doi:10.1126/science.aay0668}, such as the triangular-lattice systems  $\rm \kappa-(BEDT-TTF)_2Cu_2(CN)_3$~\cite{PhysRevLett.91.107001} and $\rm NaYbSe_2$~\cite{pratt_2011,PhysRevX.11.021044}, and the kagom\'{e} systems herbertsmithite~\cite{WOS:000312488200051} and Barlowite~\cite{Barlowite}.

\textit{Acknowledgements--}
This work was supported by the Department of Energy, Office of Science, Basic Energy Sciences under Award No. DE-SC0024469. C.B. was supported by the Gordon and Betty Moore Foundation EPIQS program under GBMF9456. A.K. and O.A.S. are supported by Grant No. 2024200 from the United States-Israel Binational Science Foundation (BSF). A.K.\ acknowledges funding by the Israel Science Foundation (Grant No.\ 2443/22). The Institute for Quantum Matter was supported through a generous donation by William H. Miller III to the Department of Physics and Astronomy. We thank the Paul Scherrer Institute for the allocated beamtime on CAMEA under proposal number 20211291. Calculations were performed using computational resources of the Maryland Advanced Research Computing Center and the Advanced Research Computing at Hopkins (ARCH) Rockfish cluster. The Rockfish cluster is supported by the National Science Foundation Award No. OAC-1920103. O.A.S. thanks Oleg Tchernyshyov and the Department of Physics and Astronomy of Johns Hopkins University for their hospitality during his extended visit in the summer of 2024, when this work was initiated.

\bibliographystyle{apsrev4-2}
\bibliography{references.bib}

\end{document}